\documentclass[runningheads]{llncs}
\usepackage[pagebackref,breaklinks,colorlinks]{hyperref}
\usepackage{orcidlink}

\usepackage{amsmath}
\usepackage{algorithm}
\usepackage{algpseudocode}
\usepackage{array}
\usepackage{booktabs}
\usepackage{caption}
\usepackage{enumitem}
\usepackage{float}
\usepackage[T1]{fontenc}
\usepackage{graphicx}
\usepackage{multirow}

\restylefloat{algorithm}

\begin{document}

\title{A More Advanced Group Polarization Measurement Approach Based on LLM-Based Agents and Graphs}

\titlerunning{A More Advanced Group Polarization Measurement Scheme}

% \author{
% Zixin Liu\inst{1}\orcidID{0009-0003-2353-7847} \and
% Ji Zhang\inst{1}\orcidID{1111-2222-3333-4444} \and
% Yiran Ding\inst{1}\orcidID{2222--3333-4444-5555}
% }

% \author{Ji Zhang$^{*}$ \orcidlink{0009-0000-1759-5184} \qquad Yiran Ding\thanks{The first two authors contribute equally.} $\ $\orcidlink{0009-0005-8624-8545} \qquad Zixin Liu$\ $\orcidlink{0009-0003-2353-7847}\\
% Wuhan University\\
% Wuhan, Hubei Province, P.R.China. 430072\\
% {\tt\small jizhang@whu.edu.cn, yrding@whu.edu.cn, liuzixin@whu.edu.cn}
% }
\author{Zixin Liu$^{*}$ \orcidlink{0009-0003-2353-7847} \qquad Ji Zhang\thanks{The first two authors contribute equally.} \orcidlink{0009-0000-1759-5184} \qquad Yiran Ding$\ $\orcidlink{0009-0005-8624-8545}\\
Wuhan University\\
Wuhan, Hubei Province, P.R.China. 430072\\
{\tt\small jizhang@whu.edu.cn, yrding@whu.edu.cn, liuzixin@whu.edu.cn}
}

\authorrunning{Zixin Liu et al.}

\institute{Wuhan University, Wuhan, Hubei 430072, China}

\maketitle

\begin{abstract}

Group polarization is an important research direction in social media content analysis, attracting many researchers to explore this field. Therefore, how to effectively measure group polarization has become a critical topic. Measuring group polarization on social media presents several challenges that have not yet been addressed by existing solutions. First, social media group polarization measurement involves processing vast amounts of text, which poses a significant challenge for information extraction. Second, social media texts often contain hard-to-understand content, including sarcasm, memes, and internet slang. Additionally, group polarization research focuses on holistic analysis, while texts is typically fragmented. To address these challenges, we designed a solution based on a multi-agent system and used a graph-structured Community Sentiment Network (CSN) to represent polarization states. Furthermore, we developed a metric called Community Opposition Index (COI) based on the CSN to quantify polarization. Finally, we tested our multi-agent system through a zero-shot stance detection task and achieved outstanding results. In summary, the proposed approach has significant value in terms of usability, accuracy, and interpretability.


\end{abstract}

\section{Introduction}

With the development of internet technology, social media has gained widespread popularity. \textit{GLOBAL DIGITAL REPORT}~\cite{ref_digi} indicate that platforms such as Facebook, YouTube, and TikTok boast billions of users worldwide. Social media has become a key avenue for the public to express opinions and engage in discussions. Its anonymity and convenience enable users to freely express their true views, thereby shaping social media public opinion. As a result, research in this field has also flourished.

In these studies, research from the perspective of group polarization holds a significant position. The concept of group polarization was first introduced by Stoner~\cite{ref_ston}, who observed that group decisions tend to be more extreme compared to individual decisions~\cite{ref_isen}. In the internet era, group polarization is broadly defined as the divergence of public opinions or stances. Building on this definition, researchers have conducted extensive and comprehensive studies on various issues related to group polarization. One of the fundamental research problems in the field of social media group polarization is its measurement. Early measurement methods based on statistical approaches~\cite{ref_bila,ref_gaur,ref_hart,ref_jaid,ref_tuma} suffered from issues such as overly simplistic for the complexity of social media dynamics. Current mainstream methods, such as text clustering or sentiment classification~\cite{ref_belc,ref_jian_2,ref_ribe,ref_tyag}, struggle to balance efficiency and accuracy. While some researchers have made significant progress by focusing on the relationships between different viewpoints~\cite{ref_boxe,ref_iyen_1,ref_jian_1,ref_lelk,ref_maia}, these studies still fall short in understanding the deeper nuances of opinion stances and their evolution.

To address the existing issues in measuring group polarization and improve efficiency, accuracy, and interpretability, we propose a new group polarization measurement approach based on LLM-based agents and graphs. This approach draws inspiration from the stance detection task in natural language processing (NLP) and the earlier "Sentiment Thermometer" method. We use a \textbf{Community Sentiment Network (CSN)} represented by a graph structure to model the polarization state, where LLM-based agents are employed to construct the network. Additionally, we design polarization measurement metrics based on CSN. To validate the effectiveness of our approach, we tested the module responsible for constructing CSN on zero-shot stance detection tasks, and the results demonstrated its superiority in capturing the nuances of group polarization.

In summary, our contributions are as follows: (1) We propose a temporal Community Sentiment Network (CSN) to represent the polarization state over time. (2) We introduce LLM-based Agents for stance detection into group polarization measurement, significantly enhancing both efficiency and accuracy. (3) We propose a more robust metric based on the CSN, Community Opposition Index (COI).

\section{Related Work}

\subsection{Opinion Polarization Measurement}

As research on group polarization deepens, extensive exploration has also been conducted on the measurement of group polarization, leading to the development of a relatively comprehensive system of measurement approaches. Existing research~\cite{ref_bila,ref_jaid} suggests that the current mainstream group polarization measurement schemes can be primarily divided into three categories: volume-based, sentiment-based, and network-based. We will discuss the characteristics of these three measurement schemes and their shortcomings when applied to the task of measuring group polarization on social media.

Volume-based approaches primarily rely on statistical methods and were widely applied in the early research of group polarization. Early researchers collected data through surveys and experiments and used statistical analysis to obtain relevant polarization results. In current trend of exploring group polarization via social media, volume-based schemes focus more on various data metrics and employ statistical methods in research. Notable examples include Gaurav et al.~\cite{ref_gaur}'s political polarization study based on the moving average aggregate probability method, Tumasjan et al.~\cite{ref_tuma}'s analysis of political polarization using the LIWC tool, and Hart et al.~\cite{ref_hart}'s use of multidimensional statistical analysis to analyze polarization during COVID-19.

However, existing studies suggest that volume-based schemes have limitations in terms of capturing information and analyzing large datasets. They fail to accurately understand the opinions and sentiments conveyed in the text and generally rely on broad statistical metrics (e.g. word frequency, likes, bookmarks, etc) to gather limited information. The lack of rapid information capture and in-depth understanding makes these techniques less effective for tracking and real-time analysis of group polarization, and they also fall short in terms of precision in measuring polarization.

Compared to volume-based approaches, sentiment-based approaches place greater emphasis on the meaning and emotions conveyed in the text. Typically, sentiment-based approaches are grounded in natural language processing (NLP) and analyze social media text from the perspectives of opinions and emotions. These methods generally follow two main strategies. The first involves clustering texts based on the similarity of sentiments, such as the IOM-NN method proposed by Belcastro~\cite{ref_belc} for accurately detecting emotional information in political polarization. The second strategy leverages deep learning for direct sentiment classification, exemplified by Tyagi et al.~\cite{ref_tyag}’s research on polarization driven by climate change, and the explorations by Ribeiro et al.~\cite{ref_ribe} and Jiang et al.~\cite{ref_jian_2} on the relationship between misinformation and polarization.

Unfortunately, both strategies face notable challenges in practice. For text clustering, current clustering algorithms are relatively coarse and simplistic, making it difficult to distinguish between disruptive information (such as advertisements or neutral statements) and significant content. Moreover, they do not account for the specific relationships between subgroups or their contribution to polarization, resulting in outcomes that lack precision and interpretability. In sentiment classification, current methods often rely on binary classifications, failing to capture the intensity of emotions. This oversimplified method negatively impacts both the interpretability and accuracy of polarization measurement.

It is also worth noting that in political polarization research, a method called the "Sentiment Thermometer" has been widely adopted~\cite{ref_boxe,ref_iyen_1,ref_iyen_2,ref_lelk,ref_wake}. This approach uses surveys to gather voters’ emotional scores toward various political parties, thus enhancing the precision and interpretability of the analysis. However, this method is costly, limited by small sample sizes, and not well-suited for measuring polarization in the context of social media.

Network-based approaches represent an approach that evaluates group polarization by considering social positions and relationships among groups, focusing on emotional direction and the stance between subgroups to better measure polarization levels. Traditional social network analysis in group polarization studies explores peripheral connections around core opinions to assess subgroup positions and emotions~\cite{ref_brav,ref_cono,ref_garc,ref_guer,ref_meda,ref_vica}. Some researchers have further developed this by dynamically simulating the process of group polarization to explore its evolutionary pathways~\cite{ref_maia,ref_sant}. With the advancement of Graph Neural Networks (GNNs), network-based approaches have been enhanced, as demonstrated by valuable explorations from researchers such as Xiao et al.~\cite{ref_xiao}, Zhang et al.~\cite{ref_zhan_3}, and Jiang et al.~\cite{ref_jian_1}, who utilized sentiment networks and GNNs in their studies. While these studies have achieved promising results in improving the scientific rigor and interpretability of group polarization research, they generally lack detailed analysis of the textual content. Moreover, social division within purely social networks does not necessarily result from group polarization, raising questions about the accuracy of some conclusions drawn from these methods.

\subsection{LLM-Based Agents}

With the introduction of OpenAI's GPT series of large language models (LLMs)~\cite{ref_brow}, numerous research fields have incorporated or examined GPT's capabilities. In the realm of group polarization research, some researchers have also explored its applications. For instance, Lu et al.~\cite{ref_luhc} used agent-based simulations to model group polarization dynamics, while Zhang et al.~\cite{ref_zhan_1} employed LLM-Based Agents to detect stances within polarized groups. These studies have yielded promising results, demonstrating the feasibility and value of applying large language models in this field.

\section{Method}

As we mentioned in Section 1, to achieve an accurate measurement of group polarization, our proposed method consists of three parts, specifically:

\begin{enumerate}[label=(\roman*)]
    \item A Community Sentiment Network (CSN) used to represent subgroups, the emotions between and within subgroups.
    \item A efficient multi-agent system for CSN construction.
    \item A group polarization metric based on the CSN, Community Opposition Index (COI).
\end{enumerate}

The set of opinion texts will be used to identify subgroups and conduct sentiment analysis through the multi-agent system, forming a CSN. The current polarization measurement result for the time slice can then be calculated using the COI.

\subsection{Community Sentiment Network}

\begin{figure}
\includegraphics[width=\textwidth]{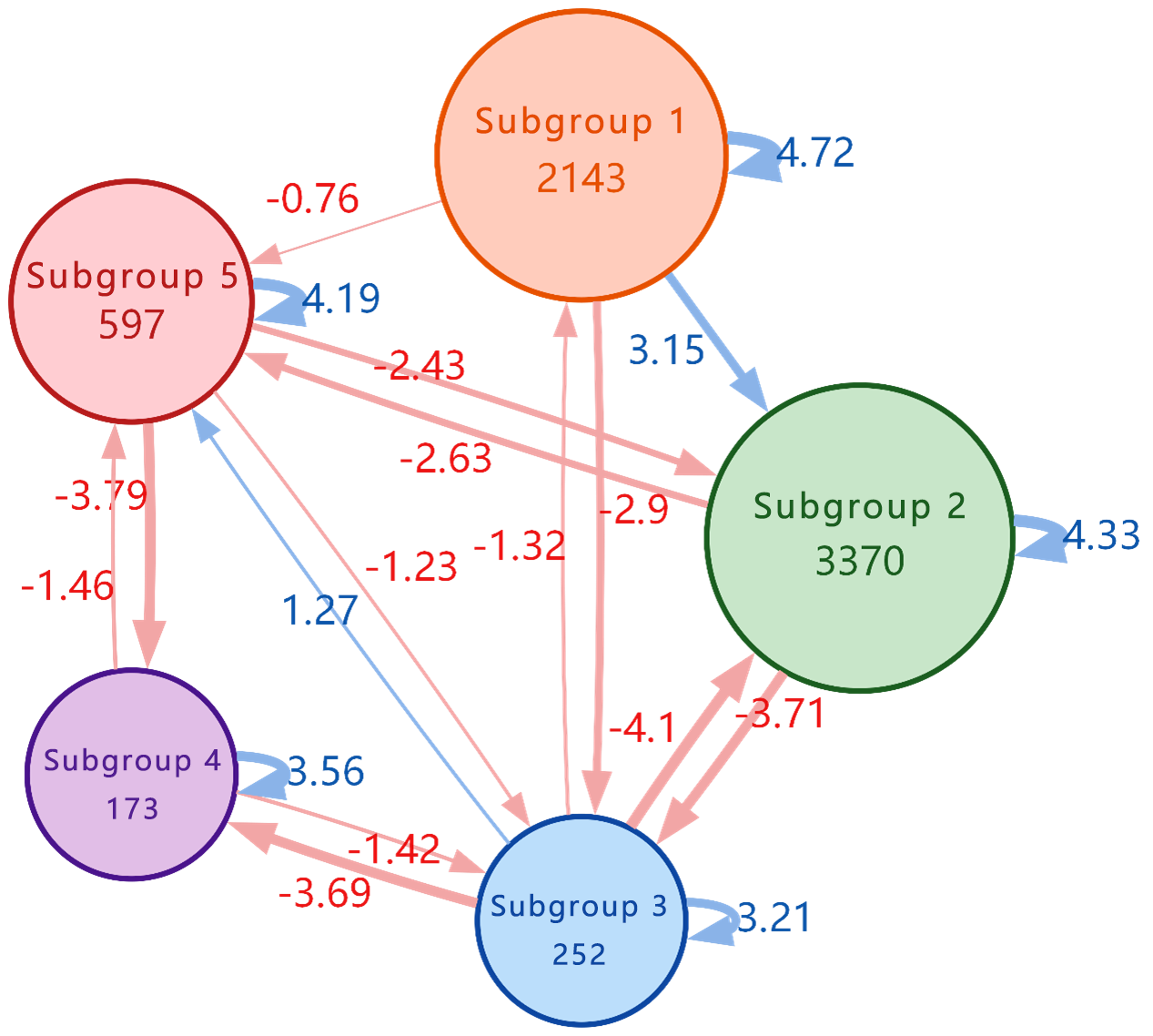}
\caption{An example of a CSN generated by Graphviz. This CSN was generated based on comments during the Russia-Ukraine conflict.} 
\label{fig_csn}
\end{figure}

The \textbf{Community Sentiment Network} (CSN, see Fig.~\ref{fig_csn}) is an extension of the "Sentiment Thermometer" method. The traditional "Sentiment Thermometer" could only be applied to two subgroups~\cite{ref_iyen_2}. CSN extends the "Sentiment Thermometer" to a directed cyclic graph that involves emotions between multiple subgroups (see Figure 1). Let $G = (V, E)$ be a graph, where $V$ is the vertex set containing subgroups and $E$ is the set of edges representing the sentimental relationships between subgroups. Each edge $e \in E$ is defined as $(u, v, s)$ where $u, v \in V$ and $s$ is the sentiment score. It should be noted that nodes $v$ and $u$ are allowed to be the same node, meaning self-loops are permitted. The score $s$ can be either positive or negative, reflecting the positive or negative nature of the sentiment.

Unlike traditional social networks, CSN uses sentiment rather than interactions as the basis for constructing connections. Also, CSN clearly illustrates the various subgroups with different stances within the target time period and reveals the emotional states between subgroups as well as the internal cohesion of each subgroup. Therefore, compared to clustering results or social networks, CSN significantly highlights the contributions of different subgroups to polarization, providing greater interpretability of the polarization state.

\subsection{A Multi-Agent System For CSN Construction}

\begin{figure}
\includegraphics[width=\textwidth]{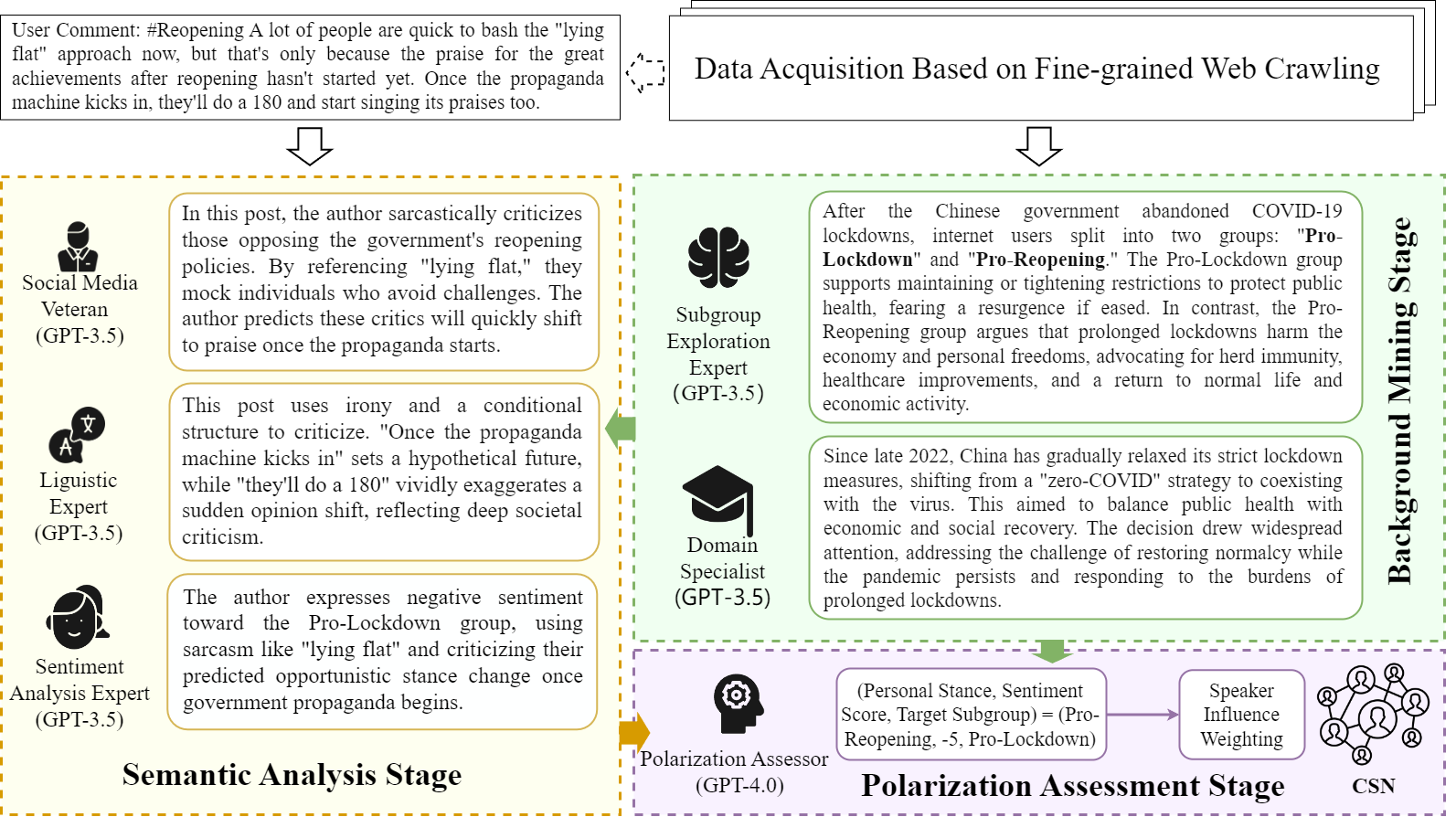}
\caption{The structure of multi-agent system for CSN construction, containing Backround Mining Stage, Semantic Analysis Stage and Polarization Assessment Stage.} \label{fig_structure}
\end{figure}

The construction of the CSN involves multiple issues, such as subgroup identification, stance detection of opinion information, and sentiment recognition. However, existing research in the field of group polarization is insufficient to provide effective solutions to these issues. Therefore, inspired by the advancements in stance detection tasks~\cite{ref_xiao}, we designed a multi-agent system based on LLM-based agents (see Fig.~\ref{fig_structure}).

In summary, our multi-agent system is composed of three stages: the Background Mining Stage, the Semantic Analysis Stage, and the Polarization Assessment Stage. The Background Mining Stage consists of Subgroup Exploration Expert and Domain Specialist, the Semantic Analysis Stage includes Social Media Veteran, Linguistic Expert, and Sentiment Analysis Expert. The Polarization Assessment Stage comprises Polarization Assessor. 
The agents collaborate through a series of interactions to provide accurate subgroup division results and reliable sentiment scores (see Alg. ~\ref{alg_1}). We will provide a detailed explanation of each stage and the functions of the respective agents in the following paragraphs. 

\begin{algorithm}
\caption{Triplets Construction Algorithm}
\label{alg_1}
\begin{algorithmic}[1]
\State \textbf{Input:} Comments data within a specified time range
\State \textbf{Output:} Set of triplets as analysis results

\Procedure{AnalyzeComments}{$comments$}
    \State $bg \gets$ \Call{DomainSpecialist}{$comments$} \Comment{bg for background}
    \State $sg \gets \emptyset$ \Comment{sg for subgroups}
    \State $uncertainComments \gets []$
    \State \Call{SubgroupExplorationExpert}{$bg$}
    \For{each $comment$ in $comments$} 
        \State $grp \gets$ \Call{SubgroupExplorationExpert}{$comment$}
        \If{not $grp$} 
            \State $uncertainComments.add(comment)$
            \If{length($uncertainComments$) reaches threshold}
                \State $grp \gets$ \Call{HumanExpertHelp}{$uncertainComments$}
                \State $sg.add(grp)$
                \State $uncertainComments \gets []$ 
            \EndIf
        \Else
            \State $sg.add(grp)$
        \EndIf
    \EndFor

    \State \textbf{Initialize Experts:}
    \State \Call{SocialMediaVeteran}{$sg, bg$}
    \State \Call{LinguisticExpert}{$bg$}
    \State \Call{SentimentAnalysisExpert}{$sg$}
    \State \Call{PolarizationAssessment}{$sg, bg$}
    \State $triplets \gets \emptyset$

    \For{each $comment$ in $comments$} 
        \State $out1 \gets$ \Call{SocialMediaVeteran}{$comment$}
        \State $out2 \gets$ \Call{LinguisticExpert}{$comment$}
        \State $out3 \gets$ \Call{SentimentAnalysisExpert}{$comment,out1,out2$}
        \State $out4 \gets$ \Call{PolarizationAssessment}{$comment,out3$}

        \If{$out4.group$}
            \State $triplet \gets (out4.group, out4.score, out4.targetGroup)$
        \Else
            \State $triplet \gets (\text{null}, out4.score, out4.targetGroup)$
        \EndIf

        \State $triplets.add(triplet)$
    \EndFor

    \State \Return $triplets$
\EndProcedure

\end{algorithmic}
\end{algorithm}

\subsubsection{Background Mining Stage} For the construction of the CSN, the first issue we need to address is how the multi-agent system understands the overall event within the context of the event. To solve this, we propose the Background Mining Stage, which uses textual information to understand the event and identify potential subgroups. Its functionality can be described as follows:

\textbf{Input:} All the comment texts related to the target topic (sampled if necessary).

\textbf{Output:} A description of the event's background, all potential subgroups present in the topic, and detailed descriptions of each subgroup.

\paragraph{Domain Specialist} Domain Specialist is primarily responsible for extracting the event background described in the comment texts. Their main tasks include exploring the core event of the topic and key stakeholders. The Domain Specialist develops a comprehensive description of the event's timeline and related parties, providing this background information to the Subgroup Exploration Expert and other subsequent stages of the process.

\paragraph{Subgroup Exploration Expert} The task of the Subgroup Exploration Expert is to use the background information provided by the Domain Specialist along with the source texts to identify potential subgroups involved in the event and summarize the possible speaking patterns of each subgroup's members. This requires the expert to explore the organizations, stances, religions, and other social identities referenced in the texts and form subgroups based on the similarity of their expressions. It is important to note that if there is a significant amount of unclassifiable content, the expert is permitted to consult human experts for clarification.

\subsubsection{Semantic Analysis Stage}
Building on the background information, Semantic Analysis Stage needs to address the primary challenge of accurately interpreting the emotions conveyed in the texts, especially when slang, homophones, sarcasm, and other nuanced expressions are present. To achieve better results on this complex task, we design the system with the specific attributes of social media in mind. Its functionality can be described as follows:

\textbf{Input:} Comment texts under the target topic and results from Background Mining Stage.

\textbf{Output:} Sentiment analysis result.

\paragraph{Social Media Veteran} The Social Media Veteran is one of the key agents responsible for semantic understanding of social media content. Its main role is to explore the patterns and characteristics of language expression on social media platforms. The agent needs to interpret the actual meanings of hashtags, internet slang, memes, and other unique forms of expression commonly used on social media. After this analysis, the Social Media Veteran passes the comprehended information to the Sentiment Analysis Expert for further processing.

\paragraph{Linguistic Expert} The Linguistic Expert is another key agent responsible for the semantic understanding of social media content. Unlike the Social Media Veteran, the Linguistic Expert focuses on analyzing the text from a linguistic perspective, examining aspects such as grammatical structure, rhetorical devices, word choice, and tense. The analysis results are also passed to the Sentiment Analysis Expert. It is important to note that the Linguistic Expert's analysis is not conducted in isolation; the background information provided by the Domain Expert supports and informs the linguistic analysis.

\paragraph{Sentiment Analysis Expert} The Sentiment Analysis Expert is the agent responsible for synthesizing various inputs to determine the final sentiment and its direction. It combines the emotional language present in the text with the semantic analysis results provided by other agents, such as the Social Media Veteran and the Linguistic Expert, to derive the sentiment of the text. Additionally, it utilizes the subgroup information provided by the subgroup exploration expert to identify the potential target of the sentiment. The results of this sentiment analysis will serve as the output of the sentiment analysis stage and will be passed to the next stage.

\subsubsection{Polarization Assessment Stage}

The primary task of the Polarization Assessment Stage is to utilize the information from the Background Mining Stage and the Semantic Analysis Stage to generate CSN in the form of triplets. Its functionality can be described as follows:

\textbf{Input:} All information from the Background Mining Stage (including potential subgroups and event background), each text from the target topic, and their sentiment analysis results.

\textbf{Output:} Sentiment expressed in the form of triplets for each comment: (personal stance, sentiment score, target subgroup).

\paragraph{Polarization Assessor} The Polarization Assessor is the core agent of the Polarization Assessment Stage. It is responsible for analyzing the author's stance, sentiment score, and the target group of the sentiment for each comment, based on the information passed from the other stages. The Polarization Assessor integrates this information into triplets. With the multi-dimensional and in-depth analysis provided by the other stages, the Polarization Assessor can make precise judgments and give credible sentiment score.

\paragraph{CSN Construction} To construct the final Community Sentiment Network (CSN) based on the sentiment triplets from the existing comments, we designed the relevant algorithm (see Alg.~\ref{alg_2}, Table~\ref{table_1} explains some variables). We use an adjacency matrix, $adjMatrix$, to represent the CSN, where $adjMatrix[i][j]$ represents the sentiment score of subgroup $i$ towards subgroup $j$. We first use all the triplets to build an initial network and then merge the nodes that belong to the same subgroup. During the merging process, we use the number of likes on the comments as a weighting factor, applying a weighted calculation to the sentiment scores between subgroups involved in the sentiment. This results in a total sentiment score between the subgroups. Since not all triplets have a personal stance, we complete them by approximating the occurrence frequency of all known personal stances as probabilities and use these probabilities to fill in the incomplete triplets. These operations results in the final CSN.

\begin{table}[h]
\centering
\caption{The explanation of some variables in Alg.~\ref{alg_2}.}
\label{table_1}
\begin{tabular}{c l}
\toprule
\textbf{Variable} & \multicolumn{1}{c}{\textbf{Explanation}} \\
\midrule
$weightSum$     & triplets' weight  \\
$countMatrix[i][j]$     & number of triplets whose person stance is $i$ and target subgroup is $j$  \\
$incompleteTriplets$     & variable for storing incomplete triplets  \\
$commentCount$     & number of comments in every subgroup  \\
\bottomrule
\end{tabular}
\end{table}

\begin{algorithm}
\caption{Construction of Community Sentiment Network}
\label{alg_2}
\begin{algorithmic}[1]
\State \textbf{Initialize:}
\State Initialize 10x10 matrices $adjMatrix$,  $weightSum$ and $countMatrix$ to zero
\State Initialize 1x10 vector $commentCount$ to zero
\State $incompleteTriplets \gets []$

\State \textbf{Process Complete Triplets:}
\For{each $triplet$ in $triplets$}
    \If{$triplet.group$ is not null}
        \State $src \gets triplet.group$
        \State $tgt \gets triplet.targetGroup$
        \State $weightedScore \gets triplet.score \times \max(triplet.likes, 1)$ 
        \State $adjMatrix[src][tgt] \gets adjMatrix[src][tgt] + weightedScore$
        \State $weightSum[src][tgt] \gets weightSum[src][tgt] + \max(triplet.likes, 1)$
        \State $countMatrix[src][tgt] \gets countMatrix[src][tgt] + 1$
        \State $commentCount[src] \gets commentCount[src] + 1$
    \Else
        \State $incompleteTriplets.add(triplet)$
    \EndIf
\EndFor

\State \textbf{Process Incomplete Triplets:}
\For{each $triplet$ in $incompleteTriplets$}
    \State $tgt \gets triplet.targetGroup$
    
    \State $probabilities \gets []$
    \For{$i \gets 1$ to $10$}
        \State $probabilities[i] \gets countMatrix[i][tgt] / \sum_{j=1}^{10} countMatrix[j][tgt]$
    \EndFor
    \State $src \gets$ sample a group based on $probabilities$
    \State $weightedScore \gets triplet.score \times \max(triplet.likes, 1)$
    \State $adjMatrix[src][tgt] \gets adjMatrix[src][tgt] + weightedScore$
    \State $weightSum[src][tgt] \gets weightSum[src][tgt] + \max(triplet.likes, 1)$
    \State $countMatrix[src][tgt] \gets countMatrix[src][tgt] + 1$
    \State $commentCount[src] \gets commentCount[src] + 1$
\EndFor

\State \textbf{Compute Averages:}
\For{$i \gets 1$ to $10$}
    \For{$j \gets 1$ to $10$}
        \If{$weightSum[i][j] > 0$}
            \State $adjMatrix[i][j] \gets adjMatrix[i][j] / weightSum[i][j]$
        \EndIf
    \EndFor
\EndFor

\State \textbf{Output:} Draw the group affect network using $adjMatrix$
\end{algorithmic}
\end{algorithm}

\subsection{Community Opposition Index (COI)}

We have designed a dedicated group polarization metric for CSN to derive a comparable and interpretable polarization index from its complex graph structure. In previous research on sentiment-based polarization measurement, a widely accepted viewpoint is that the stronger the internal cohesion within subgroups and the greater the hostility between subgroups, the higher the level of polarization~\cite{ref_iyen_2,ref_iyen_3,ref_lelk,ref_wake,ref_yarc}. The "Sentiment Thermometer" was developed based on this perspective, and its approach of calculating sentiment temperature differences has gained widespread recognition and practical use~\cite{ref_boxe,ref_iyen_1,ref_iyen_2,ref_lelk,ref_wake}. However, as we mentioned earlier, this method is only applicable when there are exactly two subgroups involved in group polarization. Therefore, we extended this calculation method to the multi-group domain and proposed the \textbf{Community Opposition Index (COI)}.

Firstly, we calculate the sentiment score of a subgroup towards the other subgroups:

\begin{equation}
(-e_{ij})\cdot1_{e_{ij}\le0}
\end{equation}

Here, $e_{ij}$ represents the sentiment score of subgroup $i$ towards subgroup $j$. $1_{e_{ij}\le0}$ means that we consider friendly subgroups as not contributing to the overall group polarization.

Subsequently, we sum the sentiment scores of subgroup $i$ towards all other related subgroups and take into account the internal cohesion within each subgroup.Therefore, we get the polarization score of subgroup $i$. Here, $t_i$ represents the internal cohesion of the subgroup $i$.

\begin{equation}
t_i\cdot\sum_j{(-e_{ij})\cdot1_{e_{ij}\le0}}
\end{equation}

Finally, we weight the overall sentiment score of each subgroup according to its size and calculate the final polarization score.

\begin{equation}
\sum_i({\frac{n_i}{N}\cdot t_i\cdot\sum_j{(-e_{ij})\cdot1_{e_{ij}\le0}}})
\end{equation}

Here $N$ represents the total number of comments on the target topic and $n_i$ represents the number of comments of subgroup $i$ within this topic.

It is important to emphasize that, since our metric is a relative indicator, it can avoid interference in the polarization measurement results caused by differences in the number of comments. Additionally, this metric, by focusing on both the internal and external sentiments of subgroups, offers better interpretability.

\section{Zero-Shot Experiments}

Our experiments will be conducted on the multi-agent system. Since there is no established benchmark in the field of group polarization research, we have chosen to test the system using stance detection tasks, which share a similar nature. We describe the specific setup of our experiments as follows.

\subsection{Datasets}

Based on existing work in the field of stance detection~\cite{ref_auge,ref_lial,ref_lian}, we will conduct our experiments on the following three datasets:

\textbf{SEM16}~\cite{ref_moha}. This dataset includes six different targets selected from various domains, namely Donald Trump (DT), Hillary Clinton (HC), Feminist Movement (FM), Legalization of Abortion (LA), Atheism (A), and Climate Change is a Real Concern (CC). It includes three types of stances: Favor, Against, and None.

\textbf{P-Stance}~\cite{ref_liys}. This dataset includes six different targets selected from political domains, namely Donald Trump (Trump), Joe Biden (Biden), Bernie Sanders (Sanders). It includes three types of stances: Favor and Against.

\textbf{VAST}~\cite{ref_alla_1}. This dataset includes large number of varying targets, and it includes three types of stances: Pro, Con and Neutral.

The statistics of our utilized datasets are shown in Table~\ref{table_2}. Since our model's use case is almost zero-shot, we will utilize these three datasets to conduct zero-shot stance detection. We will strictly adhere to the licensing requirements of the respective datasets.

To better evaluate the model's performance, we selected appropriate metrics based on existing literature~\cite{ref_alla_2,ref_lanx,ref_liur}. For the SEM16 and P-Stance datasets, we chose $F_{avg}$, which represents the average of F1 scores for Favor and Against. For the VAST dataset, we opted for Macro-F1 as the evaluation metric.

\begin{table}[h]
\centering
\caption{Statistics of our utilized datasets.}
\label{table_2}
\begin{tabular}{l l c c c}
\toprule
\textbf{Dataset} & \textbf{Target} & \textbf{Pro} & \textbf{Con} & \textbf{Neutral} \\
\midrule
\multirow{5}{*}{\textbf{SEM16}} & DT  & 148 (20.9\%) & 299 (42.3\%) & 260 (36.8\%) \\
                                & HC  & 163 (16.6\%) & 565 (57.4\%) & 256 (26.0\%) \\
                                & FM  & 268 (28.2\%) & 511 (53.8\%) & 170 (17.9\%) \\
                                & LA  & 167 (17.9\%) & 544 (58.3\%) & 222 (23.8\%) \\
                                & A   & 124 (16.9\%) & 464 (63.3\%) & 145 (19.8\%) \\
                                & CC  & 335 (59.4\%) & 26  (4.6\%)  & 203 (36.0\%) \\
\midrule
\multirow{3}{*}{\textbf{P-Stance}} & Biden    & 3217 (44.1\%) & 4079 (55.9\%) & - \\
                                   & Sanders  & 3551 (56.1\%) & 2774 (43.9\%) & - \\
                                   & Trump    & 3663 (46.1\%) & 4290 (53.9\%) & - \\
\midrule
\textbf{VAST} & - & 6952 (37.5\%) & 7297 (39.3\%) & 4296 (23.2\%) \\
\bottomrule
\end{tabular}
\end{table}

\subsection{Model Adjustment}

Since the primary purpose of our designed multi-agent system is to construct the CSN, we need to adjust the model for the experiments. We removed the Subgroup Exploration Expert and eliminated the subgroup exploration process. Instead, we input texts with predefined target groups into the remaining five agents and obtained the output from the Polarization Assessor. The adjusted model's output only includes the sentiment score and target group, making it suitable for performing stance detection tasks.

Regarding the specific details of the model configuration, we use multiple GPT-3.5 Turbo models provided by OpenAI to serve as agents in the Background Mining Stage and Semantic Analysis Stage, while GPT-4 is employed as the Polarization Assessor. This selection was primarily based on a balance between cost and the desired final performance.

\subsection{Comparison Methods}

We compare our method with various methods in zero-shot stance detection. This includes adversarial learning method: TOAD~\cite{ref_alla_2}, contrastive learning methods: PT-HCL~\cite{ref_lian}, Bert-based techniques: TGANet~\cite{ref_alla_1} and Bert-GCN~\cite{ref_liur}, LLM-based techniques: GPT-3.5 Turbo~\cite{ref_zhan_1}, GPT-3.5 Turbo+Chainof-thought (COT)~\cite{ref_zhan_2} and COLA~\cite{ref_lanx}.

\subsection{Zero-Shot Stance Detection Results}

In Table 3, we present the performance of our method on the zero-shot stance detection task, along with a comparison to other baselines. The results demonstrate that our method exhibits excellent performance in this task, with a performance improvement of 8.4\% over the current best result on the VAST dataset. Although our method did not achieve state-of-the-art (SOTA) results across all metrics, its ability to come close to or surpass current SOTA algorithms indicates its significant value when applied to group polarization research. 

\begin{table}[h]
\centering
\caption{Comparison of our method and baselines in zero-shot stance detection task, all values are percentages. Bold refer to the best performance. * denotes our method improves the best baseline at p < 0.05 with paired t-test.}
\label{table_3}
\begin{tabular}{l*{6}{>{\centering\arraybackslash}p{0.9cm}}*{3}{>{\centering\arraybackslash}p{0.9cm}}*{1}{>{\centering\arraybackslash}p{0.9cm}}}
\toprule
\multirow{2}{*}{\textbf{Model}} & \multicolumn{6}{c}{\textbf{SEM16}} & \multicolumn{3}{c}{\textbf{P-Stance}} & \multicolumn{1}{c}{\textbf{VAST}} \\
\cmidrule(lr){2-7} \cmidrule(lr){8-10} \cmidrule(lr){11-11}
& DT & HC & FM & LA & A & CC & Trump & Biden & Sanders & All \\
\midrule
TOAD          & 49.5 & 51.2 & 54.1 & 46.2 & 46.1 & 30.9 & 53.0 & 68.4 & 62.9 & 41.0 \\
TGA Net       & 40.7 & 49.3 & 46.6 & 45.2 & 52.7 & 36.6 & -    & -    & -    & 65.7 \\
BERT-GCN      & 42.3 & 50.0 & 44.3 & 44.2 & 53.6 & 35.5 & -    & -    & -    & 68.6 \\
PT-HCL        & 50.1 & 54.5 & 54.6 & 50.9 & 56.5 & 38.9 & -    & -    & -    & 71.6 \\
GPT-3.5       & 62.5 & 68.7 & 44.7 & 51.5 & 9.1  & 31.1 & 62.9 & 80.0 & 71.5 & 62.3 \\
GPT-3.5+COT   & 63.3 & 70.9 & 47.7 & 53.4 & 13.3 & 34.0 & 63.9 & 81.2 & 73.2 & 68.9 \\
COLA   & 68.5 & 81.7 & 63,4 & 71.0 & 70.8 & 65.5 & 86.6 & 84.0 & 79.7 & 73.0 \\
Ours   & \textbf{74.4}* & \textbf{81.9} & \textbf{70.3}* & \textbf{75.8}* & \textbf{76.9}* & \textbf{70.7} & \textbf{87.9} & \textbf{83.2} & \textbf{86.2}* & \textbf{81.4}* \\
\bottomrule
\end{tabular}
\end{table}

\section{Conclusion}

In this paper, we discussed the shortcomings of current group polarization measurement approaches and proposed our multi-agent and graph-based measurement scheme. Our solution innovatively introduces a large language model-based multi-agent system into the measurement of group polarization and utilizes the Community Sentiment Network (CSN) to represent the polarization state. Additionally, we provided a metric (Community Opposition Index) for calculating polarization levels using the CSN, allowing the polarization state to be quantified. Finally, we tested our multi-agent system through a zero-shot stance detection task, and the results demonstrated its usability and value.


\begin{thebibliography}{8}

\bibitem{ref_alla_1}
Allaway, E., McKeown, K.: Zero-shot stance detection: A dataset and model using generalized topic representations. arXiv preprint arXiv:2010.03640 (2020)

\bibitem{ref_alla_2}
Allaway, E., Srikanth, M., McKeown, K.: Adversarial learning for zero-shot stance detection on social media. arXiv preprint arXiv:2105.06603 (2021).

\bibitem{ref_auge}
Augenstein, I., Rocktäschel, T., Vlachos, A., Bontcheva, K.: Stance detection with bidirectional conditional encoding. arXiv preprint arXiv:1606.05464 (2016)

\bibitem{ref_belc}
Belcastro, L., Cantini, R., Marozzo, F., Talia, D., Trunfio, P.: Learning political polarization on social media using neural networks. IEEE Access \textbf{8}, 47177–47187 (2020)

\bibitem{ref_bila}
Bilal, M., Gani, A., Marjani, M., Malik, N.: Predicting Elections: Social Media Data and Techniques. In: 2019 International Conference on Engineering and Emerging Technologies (ICEET), pp. 1-6. Springer, 2019. \doi{10.1109/CEET1.2019.8711854}

\bibitem{ref_boms}
Bomsdorf, E., Otto, C.: A new approach to the measurement of polarization for grouped data. AStA Advances in Statistical Analysis \textbf{91}, 181–196 (2007)

\bibitem{ref_boxe}
Boxell, L., Conway, J., Druckman, J.N., Gentzkow, M.: Affective polarization did not increase during the coronavirus pandemic. National Bureau of Economic Research, (2020)

\bibitem{ref_brav}
Bravo, R.B., Del Valle, M.E., Gavidia, À.R.: A multilayered analysis of polarization and leaderships in the Catalan Parliamentarians' Twitter Network. In: 2015 Fifteenth International Conference on Advances in ICT for Emerging Regions (ICTer), pp. 200–206. IEEE (2015). \doi{10.1109/ICTER.2015.7377689}

\bibitem{ref_brow}
Brown, T.B., Mann, B., Ryder, N., Subbiah, M., Kaplan, J., Dhariwal, P., Neelakantan, A., Shyam, P., Sastry, G., Askell, A., Agarwal, S., Herbert-Voss, A., Krueger, G., Henighan, T., Child, R., Ramesh, A., Ziegler, D.M., Wu, J., Winter, C., Hesse, C., Chen, M., Sigler, E., Litwin, M., Gray, S., Chess, B., Clark, J., Berner, C., McCandlish, S., Radford, A., Sutskever, I., Amodei, D.: Language models are few-shot learners. arXiv preprint arXiv:2005.14165 (2020)

\bibitem{ref_cono}
Conover, M., Ratkiewicz, J., Francisco, M., Gonçalves, B., Menczer, F., Flammini, A.: Political polarization on Twitter. In: Proceedings of the International AAAI Conference on Web and Social Media, vol. 5, no. 1, pp. 89–96 (2011). \doi{10.1609/icwsm.v5i1.14126}

\bibitem{ref_digi}
DIGITAL 2024: GLOBAL OVERVIEW REPORT, \url{https://datareportal.com/global-digital-overview}, last accessed 2024/09/14

\bibitem{ref_garc}
Garcia, D., Abisheva, A., Schweighofer, S., Serdült, U., Schweitzer, F.: Ideological and temporal components of network polarization in online political participatory media. Policy \& Internet \textbf{7}(1), 46–79 (2015)

\bibitem{ref_gaur}
Gaurav, M., Srivastava, A., Kumar, A., Miller, S.: Leveraging candidate popularity on Twitter to predict election outcome. In: Proceedings of the 7th Workshop on Social Network Mining and Analysis, article no. 7, pp. 1-8. Association for Computing Machinery, New York (2013). \doi{10.1145/2501025.2501038}

\bibitem{ref_guer}
Guerra, P., Meira Jr, W., Cardie, C., Kleinberg, R.: A measure of polarization on social media networks based on community boundaries. In: Proceedings of the International AAAI Conference on Web and Social Media, vol. 7, no. 1, pp. 215–224 (2013). \doi{10.1609/icwsm.v7i1.14421}

\bibitem{ref_hart}
Hart, P.S., Chinn, S., Soroka, S.: Politicization and polarization in COVID-19 news coverage. Science Communication \textbf{42}(5), 679–697 (2020)

\bibitem{ref_isen}
Isenberg, D.J.: Group polarization: A critical review and meta-analysis. Journal of Personality and Social Psychology 50(6), 1141–1151 (1986)

\bibitem{ref_iyen_1}
Iyengar, S., Lelkes, Y., Levendusky, M., Malhotra, N., Westwood, S.J.: The origins and consequences of affective polarization in the United States. Annual Review of Political Science \textbf{22}(1), 129–146 (2019)

\bibitem{ref_iyen_2}
Iyengar, S., Sood, G., Lelkes, Y.: Affect, not ideology: A social identity perspective on polarization. Public Opinion Quarterly \textbf{76}(3), 405–431 (2012)

\bibitem{ref_iyen_3}
Iyengar, S., Westwood, S.J.: Fear and loathing across party lines: New evidence on group polarization. American Journal of Political Science \textbf{59}(3), 690–707 (2015)

\bibitem{ref_jaid}
Jaidka, K., Ahmed, S., Skoric, M., Hilbert, M.: Predicting elections from social media: a three-country, three-method comparative study. Asian Journal of Communication \textbf{29}(3), 252–-273 (2018)

\bibitem{ref_jian_1}
Jiang, J., Ren, X., Ferrara, E.: Retweet-BERT: Political leaning detection using language features and information diffusion on social networks. In: Proceedings of the International AAAI Conference on Web and Social Media, vol. 17, pp. 459–469 (2023). \doi{10.1609/icwsm.v17i1.22160}

\bibitem{ref_jian_2}
Jiang, S., Wilson, C.: Linguistic signals under misinformation and fact-checking: Evidence from user comments on social media. Proceedings of the ACM on Human-Computer Interaction \textbf{2}(CSCW), 1–23 (2018).

\bibitem{ref_knox}
Knox, R.E., Inkster, J.A.: Postdecision dissonance at post time. Journal of Personality and Social Psychology \textbf{8}(4, Pt. 1), 319 (1968)

\bibitem{ref_koga}
Kogan, N., Wallach, M.A.: Risky-shift phenomenon in small decision-making groups: A test of the information-exchange hypothesis. Journal of Experimental Social Psychology \textbf{3}(1), 75–84 (1967)

\bibitem{ref_lanx}
Lan, X., Gao, C., Jin, D., et al.: Stance detection with collaborative role-infused LLM-based agents. arXiv preprint arXiv:2310.10467 (2023)

\bibitem{ref_lelk}
Lelkes, Y., Westwood, S.J.: The limits of partisan prejudice. The Journal of Politics \textbf{79}(2), 485–501 (2017)

\bibitem{ref_lial}
Li, A., Liang, B., Zhao, J., Zhang, B., Yang, M., Xu, R.: Stance detection on social media with background knowledge. In: Proceedings of the 2023 Conference on Empirical Methods in Natural Language Processing, pp. 15703–15717 (2023). \doi{10.18653/v1/2023.emnlp-main.972}

\bibitem{ref_lian}
Liang, B., Chen, Z., Gui, L., He, Y., Yang, M., Xu, R.: Zero-shot stance detection via contrastive learning. In: Proceedings of the ACM Web Conference 2022, pp. 2738–2747 (2022). \doi{10.1145/3485447.3511994}

\bibitem{ref_liur}
Liu, R., Lin, Z., Tan, Y., et al.: Enhancing zero-shot and few-shot stance detection with commonsense knowledge graph. In: Findings of the Association for Computational Linguistics: ACL-IJCNLP 2021, pp. 3152–3157 (2021). \doi{10.18653/v1/2021.findings-acl.278}

\bibitem{ref_liys}
Li, Y., Sosea, T., Sawant, A., Nair, A.J., Inkpen, D., Caragea, C.: P-stance: A large dataset for stance detection in political domain. In: Findings of the Association for Computational Linguistics: ACL-IJCNLP 2021, pp. 2355–2365 (2021). \doi{10.18653/v1/2021.findings-acl.208}

\bibitem{ref_luhc}
Lu, H.C., Lee, H.W.: Agents of Discord: Modeling the Impact of Political Bots on Opinion Polarization in Social Networks. Social Science Computer Review (2024). \doi{10.1177/08944393241270382}

\bibitem{ref_maia}
Maia, H.P., Ferreira, S.C., Martins, M.L.: Controversy-seeking fuels rumor-telling activity in polarized opinion networks. Chaos, Solitons \& Fractals \textbf{169}, 113287 (2023)

\bibitem{ref_meda}
Medaglia, R., Zhu, D.: Public deliberation on government-managed social media: A study on Weibo users in China. Government Information Quarterly \textbf{34}(3), 533–544 (2017)

\bibitem{ref_moha}
Mohammad, S., Kiritchenko, S., Sobhani, P., Zhu, X., Cherry, C.: SemEval-2016 Task 6: Detecting stance in tweets. In: Proceedings of the 10th International Workshop on Semantic Evaluation (SemEval-2016), pp. 31–41 (2016). \doi{10.18653/v1/S16-1003}

\bibitem{ref_ribe}
Ribeiro, M.H., Calais, P.H., Almeida, V.A., Meira Jr, W.: "Everything I disagree with is \#FakeNews": Correlating political polarization and spread of misinformation. arXiv preprint arXiv:1706.05924 (2017).

\bibitem{ref_sant}
Santos, F.P., Lelkes, Y., Levin, S.A.: Link recommendation algorithms and dynamics of polarization in online social networks. Proceedings of the National Academy of Sciences 118(50), e2102141118 (2021). \doi{10.1073/pnas.2102141118}

\bibitem{ref_ston}
Stoner, J.A.: A comparison of individual and group decisions involving risk (Doctoral dissertation, Massachusetts Institute of Technology) (1961)

\bibitem{ref_tuma}
Tumasjan, A., Sprenger, T., Sandner, P., Welpe, I.: Predicting elections with Twitter: What 140 characters reveal about political sentiment. In: Proceedings of the International AAAI Conference on Web and Social Media, vol. 4, no. 1, pp. 178–185 (2010). \doi{10.1609/icwsm.v4i1.14009}

\bibitem{ref_tyag}
Tyagi, A., Uyheng, J., Carley, K.M.: Affective Polarization in Online Climate Change Discourse on Twitter. In: 2020 IEEE/ACM International Conference on Advances in Social Networks Analysis and Mining (ASONAM), pp. 443–447. IEEE, The Hague (2020). \doi{10.1109/ASONAM49781.2020.9381419}

\bibitem{ref_vica}
Vicario, M.D., Gaito, S., Quattrociocchi, W., Zignani, M., Zollo, F.: News Consumption during the Italian Referendum: A Cross-Platform Analysis on Facebook and Twitter. In: 2017 IEEE International Conference on Data Science and Advanced Analytics (DSAA), pp. 648–657. IEEE, Tokyo (2017). \doi{10.1109/DSAA.2017.33}

\bibitem{ref_wake}
Wakefield, R.L., Wakefield, K.: The antecedents and consequences of intergroup affective polarization on social media. Information Systems Journal \textbf{33}(3), 640–668 (2023)

\bibitem{ref_xiao}
Xiao, Z., Song, W., Xu, H., Ren, Z., Sun, Y.: TIMME: Twitter ideology-detection via multi-task multi-relational embedding. In: Proceedings of the 26th ACM SIGKDD International Conference on Knowledge Discovery \& Data Mining, pp. 2258–2268 (2020). \doi{10.1145/3394486.3403275}

\bibitem{ref_yarc}
Yarchi, M., Baden, C., Kligler-Vilenchik, N.: Political polarization on the digital sphere: A cross-platform, over-time analysis of interactional, positional, and affective polarization on social media. Political Communication \textbf{38}(1-2), 98–139 (2021)

\bibitem{ref_zhan_1}
Zhang, B., Ding, D., Jing, L.: How would stance detection techniques evolve after the launch of ChatGPT? arXiv preprint arXiv:2212.14548 (2022)

\bibitem{ref_zhan_2}
Zhang, B., Fu, X., Ding, D., Huang, H., Li, Y., Jing, L.: Investigating chain-of-thought with ChatGPT for stance detection on social media. arXiv preprint arXiv:2304.03087 (2023)

\bibitem{ref_zhan_3}
Zhang, C., Song, D., Huang, C., Swami, A., Chawla, N.V.: Heterogeneous graph neural network. In: Proceedings of the 25th ACM SIGKDD International Conference on Knowledge Discovery \& Data Mining, pp. 793–803 (2019).
\doi{10.1145/3292500.3330961}

\end{thebibliography}
\end{document}